# A background free double beta decay experiment

I. Giomataris

*IRFU, Centre d'études de Saclay, 91191 Gif sur Yvette CEDEX, France*

**Abstract**
We present a new detection scheme for rejecting backgrounds in neutrino less double beta decay experiments. It relies on the detection of Cherenkov light emitted by electrons in the MeV region. The momentum threshold is tuned to reach a good discrimination between background and good events.
We consider many detector concepts and a range of target materials. The most promising is a high-pressure $^{136}$Xe emitter for which the required energy threshold is easily adjusted. Combination of this concept and a high pressure Time Projection Chamber could provide an optimal solution.
A simple and low cost effective solution is to use the Spherical Proportional Counter that provides two delayed signals from ionization and Cherenkov light.
 In solid-state double beta decay emitters, because of their higher density, the considered process is out of energy range. An alternative solution could be the development of double decay emitters with lower density by using for instance the aerogel technique.
It is surprising that a technology used for particle identification in high-energy physics becomes a powerful tool for rejecting backgrounds in such low-energy experiments.

**Introduction**

Discovery of neutrino less double beta decay ββ(0ν) is one of the top priorities in contemporary neutrino physics. Observation would require Lepton Number Violation, would show that neutrinos are Majorana particles [1] and would provide precious information on the origin of neutrino masses. This process requires that neutrino is its own anti-particle and provides the most sensitive laboratory technique for measuring the neutrino mass.
Double beta decay occurs when a nucleus is energetically forbidden to decay through single beta decay. Earlier experiments [1] are providing sensitivities down to 0.2–0.8 eV scale for the effective neutrino mass using detector masses at the few tens of kg scale while present projects aim to achieve sensitivities down to 50-100 meV by increasing detector mass at the 100 kg scale. To explore the inverse hierarchy models of neutrino masses [1] will require much higher mass at the few tons scale to reach sensitivities down to 10-20 meV.
For a detector of a few tons a background level down to $10^{-5}$ c/keV/kg/year could be needed. In order to separate the tail of the ββ(2ν) distribution from the ββ(0ν) peak, which constitute an irreducible background for the latter, the energy resolution of an experiment would be kept as low as possible of the order of 1%.
The experimental challenge will include:
- A clear peak at the ββ(0ν) energy endpoint with zero background in the region surrounding the peak i.e. no other lines in the neighborhood
- A proof of the two electron nature of the ββ(0ν) event,
- A demonstration that two electron kinematics match those of ββ(0ν)

Current experiments using solid-state emitters, such as germanium or adequate bolometer detectors [2,3], dispose a particular strength to provide a record energy resolution. An example is the Heidelberg-Moscow experiment [2] searched the ββ(0ν) decay using high-purity Ge semiconductor detectors enriched to 87% in $^{76}$Ge. This experiment set the most stringent limit on the lifetime of the ββ(0ν) process.

To accommodate the other two items the use of a lighter target such as gas is necessary to identify the vertex of the interaction and track the two decaying electrons. The NEMO detector [4], still in operation at the LSM in Modane, is combining solid target and tracking in gas. Its principle of operation is based on the use of thin foils of the double beta emitter surrounded by a tracking chamber and a calorimeter. The strength of the design is a good track reconstruction which provides a topological signature useful to discriminate signal and background. The weakness is the poor energy resolution that is ultimately limited from the fact that NEMO is not a fully active detector: fluctuations are dominated by the energy loss in the target.

A competitive double beta decay candidate is the $^{136}$Xe gas whose natural abundance is rather high (9 %). Xenon gas can be easily enriched by centrifugation methods to high concentrations of $^{136}$Xe: for example, an enrichment of 80% is being used by the EXO-200 experiment [5].

A well-designed Time Projection Chamber (TPC) filled with $^{136}$Xe is a good candidate. The Gothard TPC [6] in the 90's is a pioneering experiment, with some drawbacks due the old read-out technology, like the modest energy resolution achievable. It however demonstrated the potential of a gaseous TPC to powerfully utilize the rich topological signature of ββ(0ν) events in the gas; recognize the signature as two electrons, coming from a common vertex, to further reduce background. The most dangerous background is a gamma interacting by photoelectric or Compton effect which produces an electron of an energy in the vicinity of the end point (2480 keV).

Recent progress with the Micromegas detector [7-9], with a 2 % FWHM energy resolution achieved in the MeV energy range for Xenon pressures up to 5 bar [10], together with their recently measured radio purity [11], prove that such readout is competitive for ββ(0ν) searches. This type of readout is currently used in the CAST experiment [12] providing a record low-background level.

A second read-out scheme followed by the NEXT collaboration is the electroluminescence (EL) process that presents a potentiality to improve the energy resolution [13]. Indeed both theoretical arguments [14] and a variety of experimental measurements suggest that this process could reach an energy resolution as small as predicted by the Fano factor. However the experimental result was reached in a small laboratory set-up at lower energy and therefore must be verified at the MeV energy scale using a large prototype. A draw back is that the EL process is achieved in pure Xenon which induces a degradation of the tracking accuracy; pure Xenon exhibits large diffusion coefficient and ionization electrons drifting in the long gaseous TPC are highly diffusing before reaching the read-out plane.

In this paper we propose a new read-out concept, which can be used by future double beta decay experiments, with some modifications in their experimental set-up. The purpose is to fully reject gamma backgrounds that could generate a fake ββ(0ν) peak.

**The new concept**

Xenon Time Projection Chamber (TPC) is a fully active detector, in which high-pressure noble gas acts simultaneously as target and detector. We will consider a pressure range between 10 and 20 bars as proposed by the various projects. At such pressures the xenon refractive index at the far UV (λ=150 nm) is n=1.03 and n=1.06 respectively. These values could be extracted from Figure 1 knowing the fact that n-1 is proportional to the pressure.

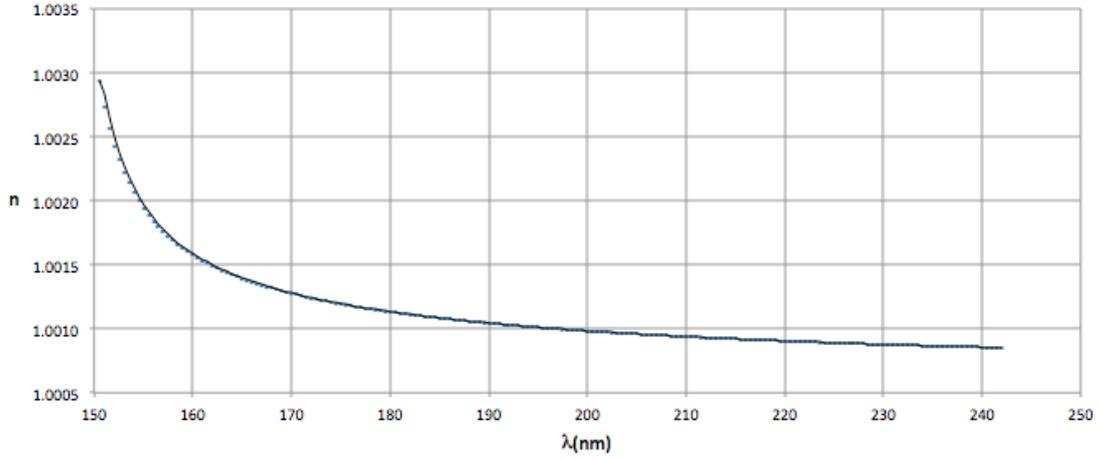

Figure 1 the Xenon refractive index as function of the light wavelength at atmospheric pressure

A ββ(0ν) event will produce two electrons sharing the total energy with a most probable energy value T=1240 keV, half of the Q value of the interaction At a pressure of 10 bar each each electron will travel a path of about 15 cm with a trajectory completely dominated by multiple scattering. The challenge is then to precisely measure a peak at the endpoint (Q=2460 keV). Backgrounds in the vicinity of this energy and especially gamma rays must be identified and rejected.
We found that this energy range is astonishing in the neighborhood of the momentum threshold required to enable Cherenkov radiation in high pressure Xenon. The main idea of our proposal is to exploit this effect in order to discriminate background electrons having energy of about 2460 keV from two electrons produced in ββ(0ν) process carrying lower kinetic energy.
Cherenkov radiation is electromagnetic radiation emitted when a charged particle (such as an electron) passes through an insulator at a constant speed greater than the speed of light in that medium.
The momentum threshold in a gaseous radiator is given by

$$p = \frac{m}{\sqrt{2(n-1)}} \qquad (1)$$

where m is the particle mass and n the refractive index of the radiator.
Using relation (1) and taking the refractive index from Figure 1 at the far UV (λ=150nm) and at pressures of 10 and 20 bar of xenon the momentum threshold for an electron is at a kinetic energy of T=2.1 MeV and T= 1.5 MeV respectively. By adjusting carefully the pressure in this range we could reach the desired result: electrons from ββ(0ν) are below momentum threshold for Cherenkov emission while background electrons will emit photons by Cherenkov radiation. Detecting these photons we will be able to recognize between two classes of electron events.
The intensity and spectrum of the radiation is given by the Frank-Tamm relation

$$\frac{dN}{dE} = \left(\frac{\alpha}{hc}\right) L(1 - 1/(n\beta)^2) \qquad (2)$$

Where dN is the number of photons with energy between E and E+dE, $\alpha$ is the fine structure constant and L the particle path in the medium.
By integrating relation (2) and taking into account the quantum efficiency QE of the photodetector we can calculate the number of detected photoelectrons $N_d$.

$$N_d = \int \frac{dN}{dE}(QE)dE \qquad (3)$$

In the approximation of a constant nb the integral over the energy bandwidth is simplified

$$N_d = N_0 L(1 - 1/(n\beta)^2) \qquad (4)$$

Where $N_d$ is the number of detected photoelectrons and $N_0$ is the Cherenkov quality factor of the detector (mean number of photoelectrons per cm):

$$No = \left(\frac{\alpha}{hc}\right) \int (QE)dE$$

From (4) we could have a first estimation of the expected signal produced by a relativistic electron ($\beta$=1) trajectory in a 10 bar Xenon gas, assuming a quality factor $N_0$ of 60 (a value easily reached by typical Cherenkov systems) and L=30 cm. Under these conditions the number of detected photons is 40, a large signal that should allow reaching full detection efficiency. A more precise estimation will be made in a next section but we would like to point out that such large yield will allow to efficiently reject background 'single' electron events having energies near the endpoint of $\beta\beta(0\nu)$. In the next sections we will give some examples how the new concept could be added to existing or future experiments.

**High-pressure Xenon TPCs**

EXO collaboration is studying two options in view of the implementation of a ton scale detector; first a liquid TPC with improved energy resolution and barium tagging and second a high-pressure gas TPC with similar features.
In case of a liquid phase Xenon the refractive index is too high (n = 1.6), therefore our concept cannot be used. The momentum threshold for Cherenkov emission is too low, therefore the pair of electrons generated from $\beta\beta(0\nu)$ process and background electrons, will emit Cherenkov light; discrimination between the two processes will not be possible. In second option, a high pressure gaseous Xenon is based on a Time Projection Chamber (TPC) with separated-function capabilities for calorimetry and tracking.
NEXT collaboration is also considering a similar high pressure TPC. Charges produced by ionization process are drifting and then collected in the end-cap readout system as used in conventional TPCs. The charge readout detector is serving for measuring the total electron energy and precisely reconstructing tracks.

In both experiments the two end caps of the TPC are covered by UV photodetectors. In order to improve energy resolution, NEXT project will be studying the possibility of reading-out light crossing the cylindrical vessel.

To incorporate our concept in these experiments we need a full 4π light collection coverage. Such system is under study by the NEXT collaboration in order to fully collect UV scintillation light produced by the electroluminescence process in pure Xenon. Under these conditions it will be difficult to separate the Cherenkov light from the very-intense scintillation light. In the case of using photodetectors sensitive only to visible and near UV light, the Cherenkov light could be possibly separated from the scintillation. A second solution is to use an adequate quencher, such as $CH_4$, mixed with Xenon gas. Such mixture will improve the gain of the detector and will vanish scintillation light emission; the light collection system will be devoted to the detection of only Cherenkov light.

**The case of the spherical detector**

The Spherical Proportional Counter (SPC) is a new type of radiation detector based on a spherical geometry [15,18]. It consists of a large spherical gas volume with a central electrode forming a radial electric field. Charges deposited in the conversion volume drift to the central sensor where they are amplified and collected. The sensor is a small spherical ball located at the center, acting as a proportional amplification structure. It allows high gas gains to be reached and operates in a wide range of gas pressures.

The main advantage of this structure is the use of a single electronic channel to read-out a large volume. This single information still allows the determination of the radial coordinate of the conversion point through the measurement of the time dispersion of the detected charge pulse. Such information is serving to localize the depth of deposits, which is a key point for applying fiducial cuts.

A good energy resolution of the order of 2 % (FWHM) is achieved in the MeV energy range as shown Figure 4.

In view of this excellent performance, that needs to be confirmed in high-pressure Xenon gas, it would make sense to consider this detector for a future ββ(0ν) investigation. Second motivation is the capability to transform this device in a full light collection photodetector.

The idea is to deposit on the internal surface a thin layer of CsI photocathode. This could be easily achieved by the vacuum deposition technology, a standard process that uses an electric resistance heater to melt the material and raise its vapor pressure to a useful range. This is done in a high vacuum, both to allow the vapor to reach the substrate without reacting with or scattering against other gas-phase atoms in the chamber, and reduce the incorporation of impurities from the residual gas in the vacuum chamber. Our spherical metallic vessel is well suited for this process with an excellent vacuum level routinely reached ($<10^{-7}$ mbar). The heater carrying the CsI sample will be installed at the center of the vessel and the photocathode will be evaporated and uniformly deposited in the whole internal surface of the vessel.

Such deposit has proven to provide a high-quantum efficiency (QE) [16-17] in the far UV bandwidth (limited by the Xenon transmission at 150nm) as shown Figure 5.

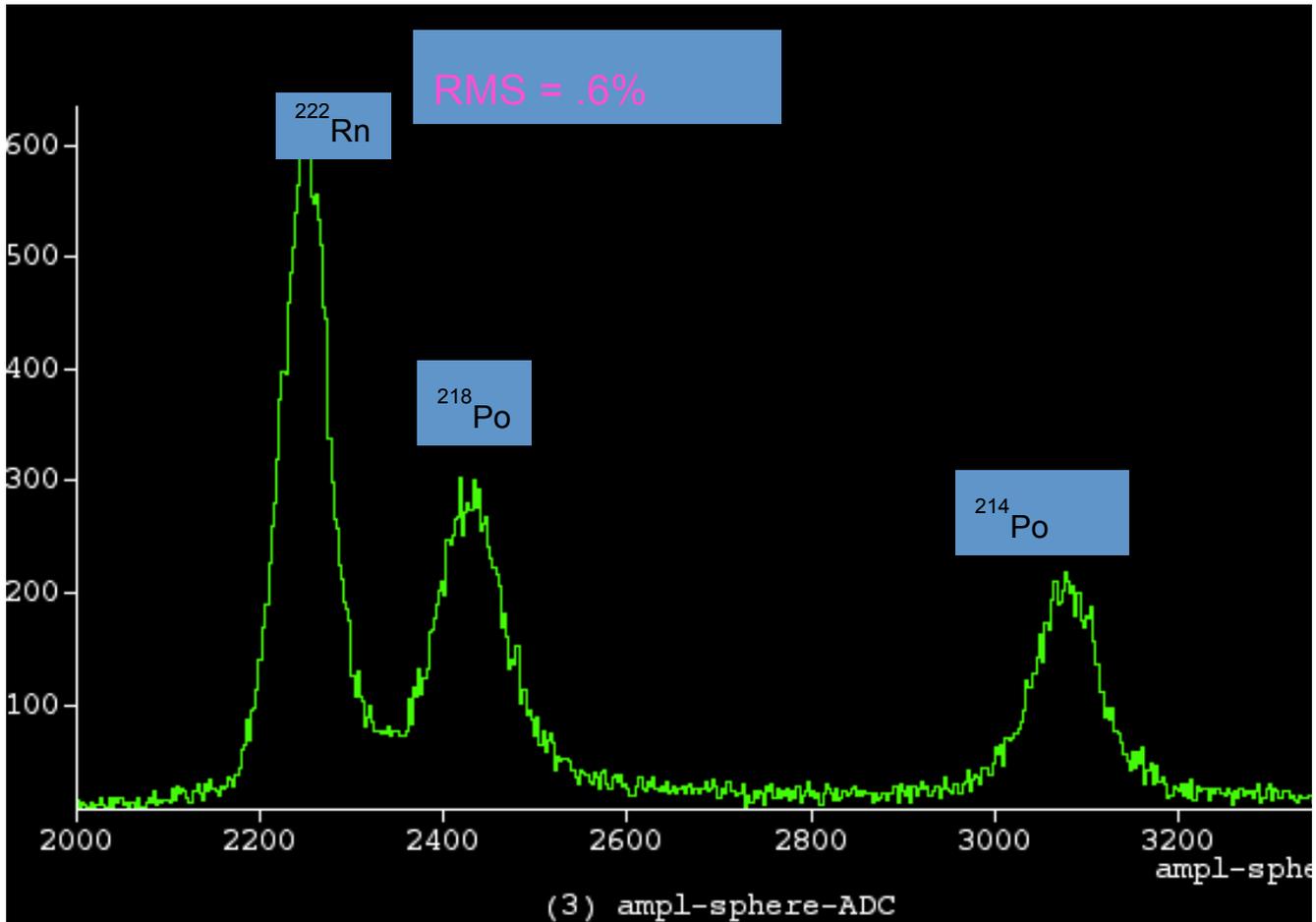

Figure 4 Peaks observed from a 222Rn radioactive source. From left to right we observe the 222Rn peak at 5.6 MeV, the 218Po and 214Po at 6.1 MeV and 7.8 MeV respectively in Argon and 2% CH4 gas mixture

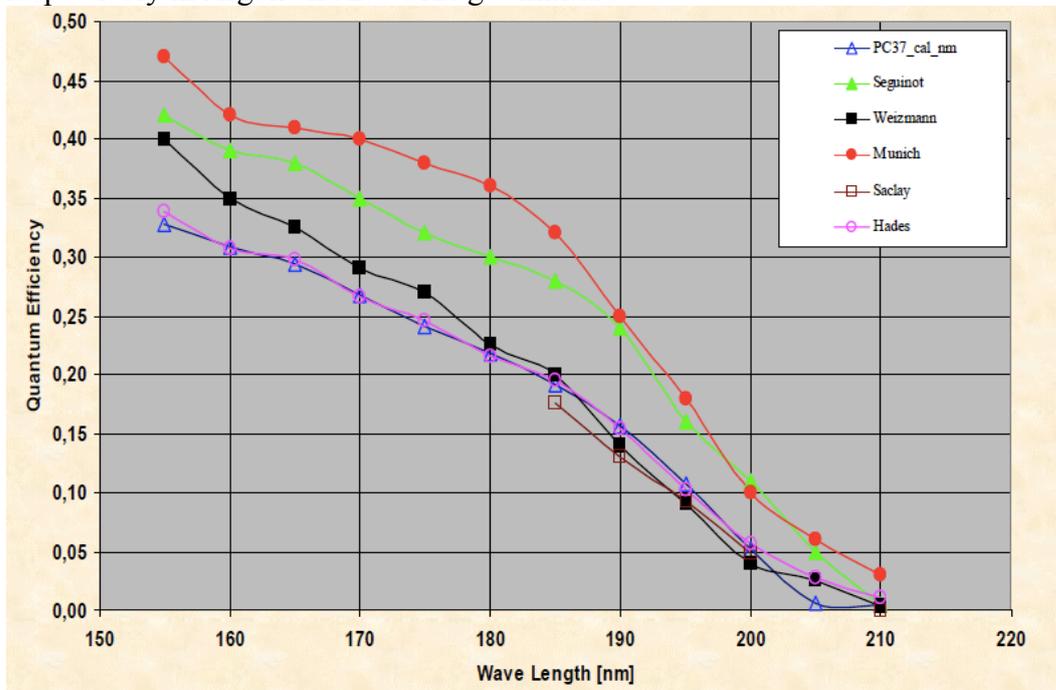

Figure 5 Quantum efficiency of CsI layer versus photon wavelength

A schematic view of the set-up is presented in Figure 6. The volume is filled with Xenon at high pressure.

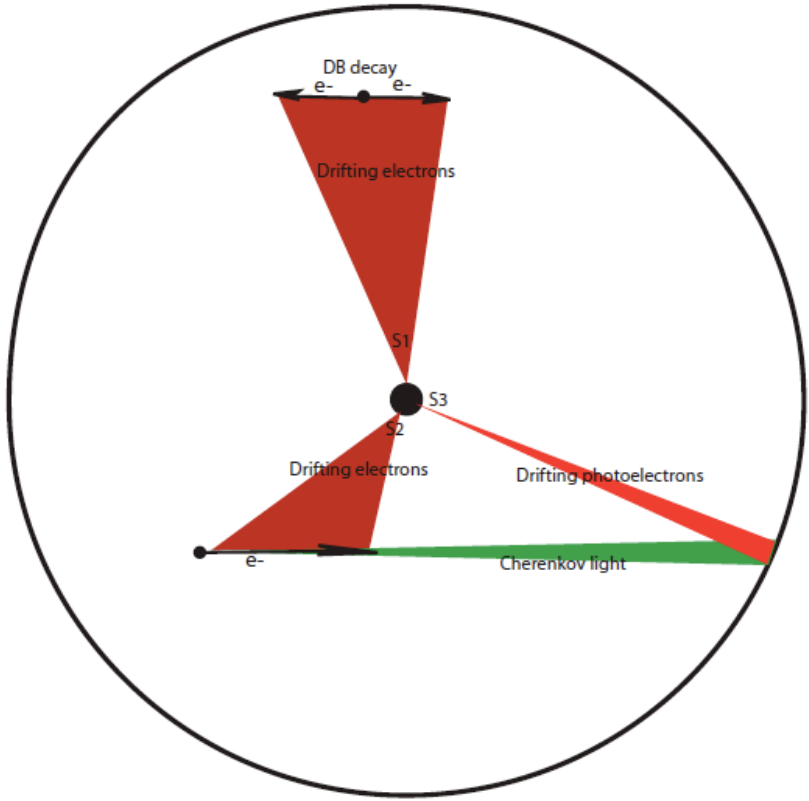

Figure 6 Two electrons from ββ(0ν) with an energy of about 1.24 MeV are ionizing the Xenon gas. Secondary electrons (in red) are drifting to the central ball where they are amplified giving rise to a signal (S1). A background electron of 2.46 MeV (above Cherenkov threshold will generate a signal (S2) by ionization process and a second signal (S3) by the Cherenkov radiation (in green) interacting with the CsI photocathode.

In this example we consider both electrons from the ββ(0ν) with an energy approximately 1.24 MeV each, below Cherenkov threshold, they will deposit their energy mostly by the ionization process. Charges are then will drift and will be collected by the central ball. The charge signal (S1) will serve for the energy measurement.
An electron of about 2.46 MeV (background at the ββ(0ν) endpoint) will deposit its energy inside the gas volume in a typical range of 30 cm. Energy loss is mainly due to ionization but at the same time UV light will be emitted by the Cherenkov process that will extract photoelectrons from the internal vessel at the CsI layer. Therefore we expect two delayed signals to be developed in the preamplifier:
- the first is a large signal (S2) due to the ionization and it is expected after a time equal to the electron drift from the primary vertex to the central sensor.
- the second due the UV Cherenkov light will induce a weaker signal (S3) at fixed delay equal to the time required for electrons to drift a radius of the sphere. The drift time has a typical value of 1 ms and depends on the size of the sphere, the gas mixture and the gas pressure.

In order to precisely calculate the yield from the Cherenkov radiation we use the Xenon refractive index from Figure 1 and the quantum efficiency of CsI from Figure

2. The efficiency for single electron detection is assumed to be 80% as it has been measured in [15]. Under these considerations we proceed to the calculation of photoelectron yield by integrating equation (3) over the available energy band width and using an electron length equal to the half of its range. Results are presented in Figure 7: the number of detected photoelectrons as a function of the electron kinetic energy for various pressures of Xenon gas. For pressures ranging from 15 bar to 28 bar the photoelectron yield the energy varies between 22 and 38 at 2.46 MeV electron energy.

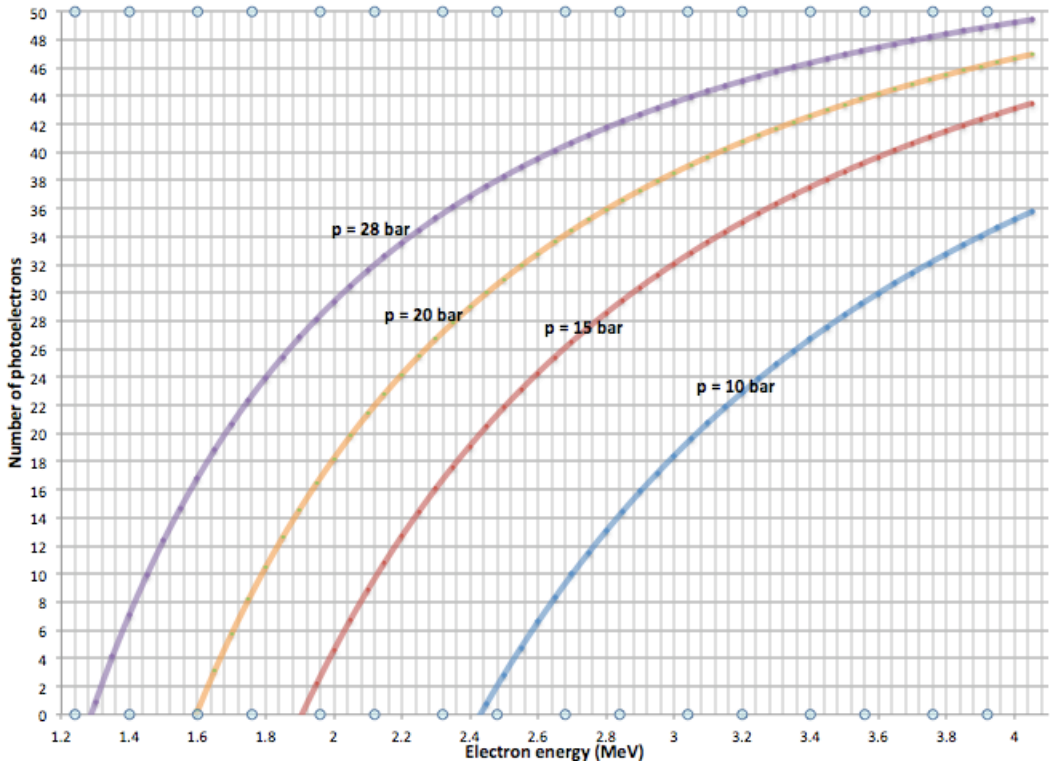

Figure 7 The number of photoelectrons as a function of the electron energy for various pressures of Xenon gas

So the delayed signal produced only by single electrons near the endpoint will be identified and rejected with a high efficiency, whereas the pair of electrons from ββ(0ν) are under the Cherenkov threshold.
To get a more precise photoelectron yield we need to measure the photoelectron extraction efficiency from the CsI under realistic conditions of $^{136}$Xenon gas at high pressure. A better estimation requires a Monte Carlo program to simulated multiple scattering and energy loss of the beta particle in the gas. Such program will be investigated in the near future.
An obvious advantage of the proposed set-up is the simultaneous reading out of charge and UV light using a single electronics channel.
A large pressurized detector volume would easily provide the required mass for the ββ(0ν) experiment. For instance the present prototype, having a diameter of 130 cm, filled with Xenon at 20 bar will contain a target mass of 137 kg. Building a larger 3m in diameter detector will provide a total mass of 1691 kg.
In summary combining the new concept with the spherical detector provide a simple and cost effective solution and the scaling up to several tons is not a big challenge. In this case we could employ a second identical system filled with raw Xenon gas running in parallel under similar conditions. Subtracting the acquired spectra from the two systems will be a way to further reject backgrounds and improve the sensitivity to observe double beta decay.

**Solid state or liquid detectors**

In the case of solid or liquid state detectors the main requirement to have a transparent medium with a refractive index of the order of 1.05 will not fulfill. Such value stands much below the range provided by solid or liquid media. Cherenkov radiation will be emitted for all electrons in the MeV energy range and therefore the discrimination between beta decay electrons and background electrons is not possible. However this signal could serve to reject heavier particles such as alphas, protons, heavy recoils. We could argue that, because of lack of adequate refractive index, solid targets are not good candidates unless we could manage to reduce their density using for instance aerogel technology. Aerogel is a manufactured material with the lowest bulk density of any known porous solid. It is derived from a gel in which the liquid component of the gel has been replaced with a gas. The result is an extremely low-density solid with several remarkable properties, transparent in visible light and sometimes in the near UV light where the transmission is determined by the light scattering length.
Studies on production of aerogel with refractive index in the range of interest, for use as the Cherenkov light radiator in RICH detectors, have being carried out and such products have been successfully developed by the industry. Silica aerogel having a colloidal form of $SiO_2$ was usually employed as bulk material. The deal will be to get them with other elements that are double beta emitters. More difficult is to preserve the good energy resolution, which could not be provided by the weak Cherenkov process. In the case of a bolometer, for instance, the accurate energy measurement should be performed through the heat measurement. In the case of a semiconductor, after the aerogel fabrication, it will be certainly difficult to preserve the semiconductor properties.
In NEMO like experiments because the target material is not active it could be replaced by an aerogel without affecting the detection scheme.
A second option is to keep the solid or liquid material as it is and proceed to a slightly different read-out approach as it is illustrated in Figure 7.

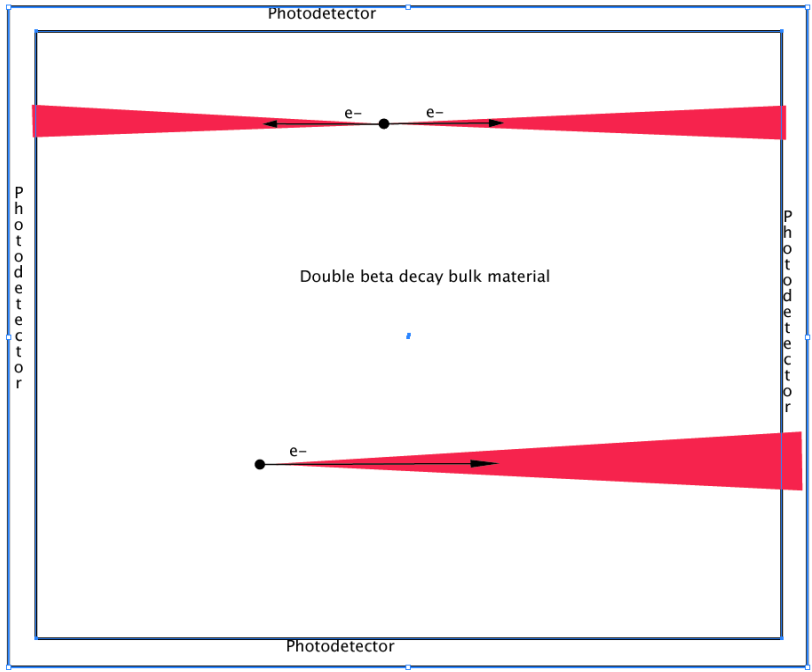

Figure 7. The internal box is the double beta emitter surrounded by photodetectors. The process at the top of the box illustrates the Cherenkov light emission by two back-to-back electrons from ββ(0ν) while at the bottom a single background electron generates a single light spot on the photodetector.

The double beta decay emitter is presented as a box surrounded by a photodetector system. In principle the device could be any of existing solid of liquid detector system. The only requirement is to be transparent in order to transmit Cherenkov light.
Now the two electrons from the ββ(0ν) process will produce Cherenkov light as shown in the figure (top of the box) because of the high density, high refractive index of the medium. Therefore we expect two back-to-back detected spots in the photodetector. This is another way of differentiating two electrons signal from a single background electron producing a single spot at the photodetector (bottom of the box). Notice the two well separated spots are an illustration of the process. Because of multiple scattering of the electron in the dense medium the presented spot will be highly diluted. This configuration needs a careful study with an appropriate simulation program.
Among *ββ* decay isotopes $^{48}$Ca has the highest Q value (4.27 MeV) and therefore low background rate from natural radioactivity is expected. The CANDLES experiment [19] is using this isotope in form of $CaF_2$ single crystal doped with Eu in order to enhance scintillation process.
$CaF_2$ is a crystal with good transmission from 130 nm to 780 nm. The available bandwidth is large (=7 eV) to allow a large amount of Cherenkov UV light to be produced by relativistic electrons. By reading out both scintillation and Cherenkov light an additional background rejection could be achieved. The drawback of this scintillator is its poor energy resolution (CANDLES is reaching 9.1% (FWHM) at 662 keV).
An idea could be to use the $CaF_2$ crystal before Eu doping and to read-out heat released by particle energy loss in the crystal employing an adequate bolometer. In this way we hope to preserve the good energy provided by the bolometer and the good background rejection offered by the Cherenkov process.

A similar idea was studied in the case of the TeO$_2$ target where Cherenkov emission from beta rays in bolometric crystal and has been proposed as a viable alternative to scintillation [20]. In the case of a bolometer the difficulty consists in the need to reach very low temperatures, close to absolute zero, and therefore adequate photodetectors capable to operate at such temperature are required.

In principle any double beta emitter could be compatible with our concept under the condition to provide good light transmission in the visible and most likely in the UV region.

A drawback of solid and liquid beta emitters is their high density that enables two side effects:
- The electron path is reduced inversely proportional to the density while the number of Cherenkov photons is reaching saturation
- The effect of electron multiple scattering is highly increased

The Xenon gas is again the most attractive candidate. By adjusting gas, as shown in previous examples, we could manage to get Cherenkov light emission from both ββ(0ν) electrons and background electrons. Figure 8 shows the calculated Cherenkov photon yield at a pressure of P= 50 bar as function of the electron energy assuming CsI UV photocathode. For background electron at 2.46 MeV energy we expect a spot at the photodetector with about 60 detected photoelectrons. Notice that in case of UV phomultiplier or adequate silicon multiplier, the available bandwidth is much larger and the photoelectron yield is expected to be even higher. The two electrons produced by the ββ(0ν) process will generate, most of the time, two photoelectron spots and therefore they will be well discriminated from single background electrons.

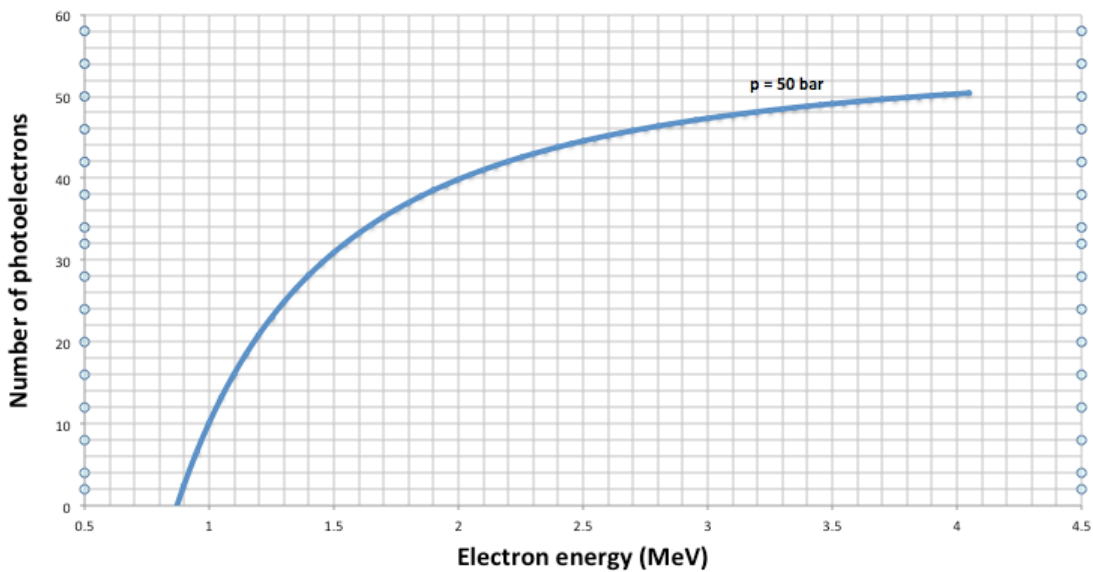

Figure 3. Number of photoelectrons as a function of the electron energy

**Conclusion and outlook**

A novel experimental approach, for eliminating gamma backgrounds in future experiments, is proposed. The new approach rely on the detection of Cherenkov light emitted by electrons with energy close to the endpoint of the ββ(0ν) process to sort between various processes.

This is an additional signature to discriminate between signal and backgrounds, which is independent and will be added to other identification cuts such as event topology or daughter ion tagging. A powerful rejection of gamma backgrounds is expected.
This scheme is well adapted to high-pressure $^{136}$Xe target which is UV transparent Cherenkov radiator and at the same time the optimal momentum threshold could be easily tuned by choosing the right gas pressure.
Because of the high refractive index, the concept is useless in the case of liquid or solid detectors but it could serve to reject heavy particles. A possible issue could be the fabrication of low-density double beta decay emitters using for instance aerogel technology.
In a second approach, the radiator has higher index and a background electron will produce a single spot by Cherenkov process different from the two electrons pattern generated by ββ(0ν) producing two light spots. In this scheme most of solid and liquid double emitters are candidates under the condition to be transparent for visible and eventually UV light. Light density emitters are preferable and in that respect $^{136}$Xe at a pressure of about 50 bar is the most promising candidate.

**REFERENCES**

[1] F.T. Avignone, III, S.R. Elliott and J. Engel, Double Beta Decay, Majorana Neutrinos and Neutrino Mass, Rev. Mod. Phys. 80 (2008) 481
[2] H. V. Klapdor-Kleingrothaus et al., Eur. Phys. J. A 12 (2001)147.
[3] C. Arnaboldi *et al.*,Phys. Rev. Lett. 95 (2005),142501.
[4] R. Arnold et al., Nucl.Instrum.Meth.A536:79-122,2005.
[5] Akimov D. et al., Nucl. Phys. Proc. Suppl. 138 (2005), 224.
[6] R. Luscher et al., Phys. Lett. B 434 (1998) 407
[7] Y. Giomataris, P. Rebourgeard, J.P. Robert and G. Charpak,
Nucl. Instrum. Meth. A 376 (1996) 29.
[8] Y. Giomataris, Nucl. Instrum. Meth. A 419 (1998) 239
[9] I. Giomataris et al., Nucl. Instrum. Meth. A 560 (2006) 405
[10] S. Cebrian et al., JCAP 1010:010,2010.
[11] S. Cebrian et al., Radiopurity of MicrOMEGAs readout planes, arXiv:1005.2022
[12] J. Galan et al.,2010 JINST 5 P01009.
[13] The NEXT collaboration, F. Granena et al., NEXT, a HPGXe TPC for neutrinoless double beta decay searches, arXiv:0907.4054
[14] D. Nygren, Nucl. Instrum. Meth. A 603 (2009) 337
[15] I. Giomataris et al., JINST 3: P090007 (2008).
[16] J. Sequinot et al., , Nucl. Instrum. Meth.A297(1990)133-147.
[17] A. Bream et al., Nucl.Instrum.Meth.A343:163-172,1994.
[18] E. Bougamont et al., arXiv:1010.4388.
[19] S. Umehara *et al.*, AIP Conf.Proc.1235:287-293,2010.
[20] T. Tabarelli de Fatis, Eur. Phys. J. C (2010) 65: 359–361